\documentclass[
twocolumn,
pra,showpacs,preprintnumbers,amsmath,amssymb]{revtex4}

\usepackage{graphicx}
\usepackage{amssymb}
\usepackage{amsmath}
\usepackage{dsfont}
\usepackage{bm}
\usepackage{mathrsfs}

\begin{document}


\title{Reply to the Comment on Perfect imaging with positive refraction in three dimensions}
\author{Ulf Leonhardt and Thomas G. Philbin}
\affiliation{
University of St Andrews, School of Physics and Astronomy,
North Haugh, St Andrews KY16 9SS, UK
}
\date{\today}
\begin{abstract}
Exact time-dependent solutions of Maxwell's equations in Maxwell's fish eye show that perfect imaging is not an artifact of a drain at the image, although a drain is required for subwavelength resolution.  
\end{abstract}
\maketitle

Perfect imaging with positive refraction \cite{LeoPerfect,LP,BMG} challenges some of the accepted wisdom of subwavelength imaging \cite{Blaikie,LeoReply,Guenneau,Merlin}. In particular, it requires a drain for perfect resolution \cite{deRosny}. Merlin \cite{Merlin} argues that the perfect focusing is an artifact of the drain. However, instead of discussing Maxwell's fish eye, he considers the focusing of light in a spherical mirror. The mirror serves as a simple model that resembles the fish eye, but this model is too simple: Maxwell's fish eye has imaging properties different from mirrors \cite{Mirror}. Furthermore, his reasoning is in conflict with experimental evidence \cite{Ma}. Let us briefly explain how perfect imaging is achieved in Maxwell's fish eye and what the role of the drain is. Further details can be found in Ref.\ \cite{Mirror}.

In Ref.\ \cite{LP} we solved Maxwell's equations for electromagnetic radiation in Maxwell's fish eye (with $\varepsilon=\mu=n$) by reducing the problem to the propagation of a scalar wave $D$; the electromagnetic field can be calculated from $D$ by certain differentiations \cite{Dover}. We thus only need to discuss the imaging properties of $D$, which does not constitute a simple model, but an exact representation of the electromagnetic Green tensor \cite{LP,Dover} in Maxwell's fish eye. 

The fish eye is characterized in terms of a length $R$ that defines the scale of the refractive index profile. For simplicity of notation, we measure space in units of $R$ and time in units of $R/c$; in these units the speed of light in vacuum is $1$. In our units the index profile of Maxwell's fish eye is given by
\begin{equation}
\label{max}
n = \frac{2}{1+r^2} \,.
\end{equation}
The scalar wave $D$ satisfies the equation \cite{LP}
\begin{equation}
\label{wave}
\frac{1}{n^3}\,\nabla\cdot n \nabla D - D - \frac{\partial^2 D}{\partial t^2} = -\frac{\delta(\bm{r}-\bm{r}_0)\delta(t)}{n^3}
\end{equation}
where we include a source term on the right-hand side that corresponds to a point source at position $\bm{r}_0$ acting in one moment of time that we set to $t_0=0$ without loss of generality. The wave equation thus describes a flash of light emitted at an arbitrary position and we can shift the time of emission to an arbitrary $t_0$. Note that any light field can be thought of as a continuous superposition of such light flashes, and so our case is sufficiently general. 

Consider the Fourier transformation of the light flash 
\begin{equation}
\label{invfourier}
\widetilde{D}(\bm{r},\omega) = \frac{1}{2\pi} \int_{-\infty}^{+\infty} D(\bm{r},t) e^{i\omega t} \, dt \,.
\end{equation}
Note that we can read the Fourier integral (\ref{invfourier}) as the amplitude of the wave
\begin{equation}
\label{invfourier2}
\widetilde{D}(\bm{r},\omega)e^{-i\omega t} = \frac{1}{2\pi} \int_{-\infty}^{+\infty} D(\bm{r},t-t_0) e^{-i\omega t_0} \, dt_0 
\end{equation}
that is created by the continuous emission of light flashes at times $t_0$ with phases $\omega t_0$. The Fourier amplitude $\widetilde{D}$ thus plays a double role: it describes the frequency components of a single flash of light emitted at $\bm{r}_0$ but also the amplitude of a stationary wave generated at the source point $\bm{r}_0$ with frequency $\omega$.

To proceed, we write the wave equation (\ref{wave}) in the frequency domain
\begin{equation}
\label{fishfourier}
\frac{1}{n^3}\,\nabla\cdot n \nabla \widetilde{D} + (\omega^2-1) \widetilde{D} = -\frac{\delta(\bm{r}-\bm{r}_0)}{2\pi n^3}
\end{equation}
that has the solution \cite{Mirror}
\begin{equation}
\label{dfish}
\widetilde{D} =  \left(r'+\frac{1}{r'}\right) \frac{\sin(2\omega\,\mathrm{arccot}\, r')}{(4\pi)^2\sin(\pi\omega)}
\end{equation}
in terms of the M\"obius-transformed radius \cite{LP,Dover}
\begin{equation}
\label{moebius}
r' = \frac{|\bm{r}-\bm{r}_0|}{\sqrt{1+2\bm{r}\cdot\bm{r}_0 + |\bm{r}|^2 |\bm{r}_0|^2}} \,.
\end{equation}
We see that the source point $\bm{r}_0$ corresponds to $r'=0$. We also see that the image point of light rays in Maxwell's fish eye \cite{Dover},
\begin{equation}
\bm{r}_0' = -\frac{\bm{r}_0}{|\bm{r}_0|^2} \,,
\end{equation}
corresponds to $r'=\infty$. The M\"obius-transformed radius thus conveniently describes the imaging in Maxwell's fish eye. However, we still need to investigate whether the Fourier amplitude (\ref{dfish}) constitutes the wave of the light flash. Causality requires that the Fourier amplitude $\widetilde{D}$ is analytic on the upper half complex frequency plane. The solution (\ref{dfish}) is decaying for $\Im\omega\rightarrow\pm\infty$. It has singularities at $\omega=m$ with integer $m$ that we move by an infinitesimal amount below the real frequency axis \cite{Mirror} such that $\widetilde{D}(\omega)$ is analytic on the upper half plane, as required. The solution (\ref{dfish}) is the correct Fourier amplitude of the light flash.

Let us adopt the alternative interpretation of the Fourier integral (\ref{invfourier}) where we regard $\widetilde{D}$ as the amplitude of a stationary wave generated at the source point $\bm{r}_0$ with frequency $\omega$. The wave (\ref{dfish}) has the real wave function of a typical standing wave --- radiation emitted from the source is reflected in Maxwell's fish eye and reabsorbed at the source, forming a stationary standing wave \cite{LeoReply}. Near the image point the continuous wave $\widetilde{D}$ behaves as
\begin{equation}
\label{spot}
\widetilde{D} \sim \frac{\omega\,\mathrm{sinc}(2\omega/r')}{8\pi^2 \sin(\pi\omega)} \quad \text{with} \quad \mathrm{sinc}\, x = \frac{\sin x}{x} \,,
\end{equation}
i.e.\ as a diffraction-limited spot. Clearly, the stationary wave is not perfectly imaged, in agreement with Merlin's Comment \cite{Merlin}. 

Now consider the time-dependent wave $D$, the flash of light emitted at position $\bm{r}_0$. We obtain from Cauchy's theorem \cite{Mirror}
\begin{eqnarray}
D &=& \int_{-\infty}^{+\infty} \widetilde{D} e^{-i\omega t} \, d\omega 
\nonumber\\
&=& \frac{\Theta(t)}{8\pi}\left(r'+\frac{1}{r'}\right) \sum_{m=-\infty}^{+\infty} \big[\delta(t-2\arctan r' - 2m\pi) 
\nonumber\\
&&
\quad\quad\quad\quad\quad\quad - \delta(t+2\arctan r' - 2m\pi)\big] \,.
\label{dflashes}
\end{eqnarray}
Equation (\ref{dflashes}) explicitly describes the time evolution of the flash: after emission at $t=0$ the light wave expands from the source point and then contracts towards the image where it focuses in a single point and is reflected. Upon reflection the wave changes sign; it expands again and then contracts towards the source point where it is reflected, changes sign, and so forth. In contrast, in Merlin's mirror the wave is distorted after the first reflection \cite{Mirror}. In Maxwell's fish eye, an individual light flash from the source point is focused at the image in a perfect point, but then the flash is reflected and changes sign. In the stationary wave (\ref{invfourier2}) a continuous stream of flashes with phases $\omega t_0$ is averaged over time. The sign change upon reflection results in the reduced resolution (\ref{spot}) that conforms to the standard diffraction limit.

The explicit solution (\ref{dflashes}) suggests an easy remedy for the imperfection in imaging with otherwise perfect lenses: give the wave an outlet at the image such that it is not reflected there \cite{Ma}. In this case the series (\ref{dflashes}) of reflections is truncated at the first term:
\begin{equation}
\label{dflash}
D= \frac{\Theta(t)}{8\pi}\left(r'+\frac{1}{r'}\right) \delta(t-2\arctan r')\,\Theta(\pi-t) \,.
\end{equation}
The Fourier transform (\ref{invfourier}) of the wave (\ref{dflash}) is  
\begin{equation}
\label{drun}
\widetilde{D} = \frac{1}{(4\pi)^2}\left(r'+\frac{1}{r'}\right) \exp(2\mathrm{i}\omega\arctan r') \,.
\end{equation}
Formula (\ref{drun}) describes a running wave with complex wave function propagating from the source to the image where the wave disappears \cite{LeoReply}, in contrast to the standing wave (\ref{dfish}) with real wave function that is reflected at the image. The spatial singularity of $\widetilde{D}$ at $r'=\infty$ corresponds to a supplementary source at the image point $\bm{r}_0'$. The outlet thus acts as a drain. As $2\arctan r'\rightarrow\pi$ for $r'\rightarrow\infty$ the wave carries a phase delay of $\pi\omega$ at the image point \cite{LP}. 

Note that the outlet at the image is completely passive \cite{deRosnyComment}. It is a point absorber with infinitely small cross section that can only extract entire waves when they are concentrated at its location with infinite intensity. An ideal point detector should play exactly this role. An array of such detectors is required for imaging. In practice, of course, detectors are not ideal and have finite cross section and finite efficiency. The cross section of detectors does indeed limit the imaging resolution \cite{Ma}, whereas the finite efficiency simply reduces the amplitude of the perfectly focused wave: the total wave becomes a superposition of the diffraction-limited wave (\ref{dfish}) and the perfectly imaged component (\ref{drun}). Maxwell's fish eye thus makes a perfect lens with point-like resolution for electromagnetic waves, but only when such waves are detected by perfect point detectors. The perfect image appears, but only if one looks.

{\it Acknowledgements.---} We are indebted to Aaron Danner, Susanne Kehr, Sahar Sahebdivan and Tom\'a\v{s} Tyc for inspiring discussions. Our work is supported by the Royal Society of Edinburgh, the Scottish government, and the Royal Society of London.


\begin{thebibliography}{99}

\bibitem{LeoPerfect}
U. Leonhardt,
New. J. Phys. {\bf 11}, 093040 (2009).

\bibitem{LP}
U. Leonhardt and T. G. Philbin, 
Phys. Rev. A {\bf 81}, 011804 (2010).

\bibitem{BMG}
P. Benitez, J.C. Mi\~{n}ano, and J.C. Gonz\'{a}lez,
Opt. Express {\bf 18}, 7650 (2010).

\bibitem{Blaikie}
R.J. Blaikie, 
New J. Phys. {\bf 12}, 058001 (2010).

\bibitem{LeoReply}
U. Leonhardt, 
New J. Phys. {\bf 12}, 058002 (2010).

\bibitem{Guenneau}
S. Guenneau, A. Diatta, and R.C. McPhedran, 
J. Mod. Optics {\bf 57}, 511 (2010).

\bibitem{Merlin}
R. Merlin, Comment on Ref.\ \cite{LP}.

\bibitem{deRosny}
J. de Rosny and M. Fink, 
Phys. Rev. Lett. {\bf 89}, 124301 (2002).

\bibitem{Mirror}
U. Leonhardt and S. Sahebdivan, 
arXiv:1007.0078.

\bibitem{Ma}
Y.G. Ma, C.K. Ong, S. Sahebdivan, T. Tyc, and U. Leonhardt,
arXiv:1007.2530.

\bibitem{Dover}
U. Leonhardt and T. G. Philbin,
{\it Geometry and Light: The Science of Invisibility}
(Dover, Mineola, 2010).

\bibitem{deRosnyComment}
This is in contrast to the perfect imaging of ultrasound by far-field time reversal \cite{deRosny} where an active drain was used that was the time reverse of the source. 

\end{thebibliography}
\end{document}